# Realization of Kagome Kondo lattice

Boqin Song[1,†], Yuyang Xie[1,2,†], Wei-Jian Li[3,†], Hui Liu[1], Jing Chen[1], Shangjie Tian[4], Hechang Lei[4], Qinghua Zhang[1], Jian-gang Guo[1], Lin Zhao[1,2,*], Shun-Li Yu[3,*], Xingjiang Zhou[1,2], Xiaolong Chen[1,2,*], Tianping Ying[1, *]

1. Beijing National Laboratory for Condensed Matter Physics, Institute of Physics, Chinese Academy of Sciences, Beijing 100190, China.
2. School of Physical Sciences, University of Chinese Academy of Sciences, Beijing 100049, China.
3. National Laboratory of Solid State Microstructures and Department of Physics, Nanjing University, Nanjing 210093, China
4. Department of Physics and Beijing Key Laboratory of Optoelectronic Functional Materials & MicroNano Devices, Renmin University of China, Beijing 100872, China

*ying@iphy.ac.cn
*lzhao@iphy.ac.cn
*slyu@nju.edu.cn
*xlchen@iphy.ac.cn

## Abstract

**The Kondo lattice, describing a grid of the local magnetic moments coupling to itinerant electrons, is a fertile ground of strongly correlated states in condensed matter physics. While the Kagome lattice has long been predicted to host Kondo physics with exotic magnetism and nontrivial topology, no experimental realization has been achieved. Here, we report the discovery of $CsCr_6Sb_6$, a van der Waals-like Kagome Kondo lattice featuring extremely flat, isolated bands at the Fermi level ($E_F$) that composed entirely of Cr-3d electrons. We observe heavy fermions with the effective mass over 100 times greater than those of its vanadium counterpart. We also observe Kondo insulating behavior in an ultra-low carrier density of $10^{19}$ $cm^{-3}$ and dimensionality-induced Kondo breakdown. More interestingly, the frustrated magnetism observed in the bulk give way to a hidden A-type antiferromagnetic ordering in few layers, in sharp contrast to the common sense of weakened magnetism with thinning. The realization of Kondo physics in Kagome lattice opens avenues for exploring diverse quantum criticalities in a strongly-correlated frustrated system.**

Kagome lattice, characterized by symmetry-protected Dirac points, van Hove singularities and flat bands, provides an exciting platform to intertwine nontrivial topology, frustrated magnetism and strongly correlated electrons [1-16]. However, accessing strong correlation effects in Kagome materials remains challenging, as their flat bands are often distant from $E_F$ or disrupted by structural distortions that deviate from the ideal Kagome structure [12-16]. Recent observations of strange metallicity in $Ni_3In$ and pressurized $CsCr_3Sb_5$ suggest that bringing portions of the Kagome flat bands closer to $E_F$ could effectively enhance electronic correlations [17-19]. Moreover, if local magnetic moments are introduced—commonly attained through strong Coulomb interactions in *f*-electron compounds [20-31]—the long-sought-after Kagome Kondo lattice could be realized (Fig. 1(a)). Local magnetic moments have been reported in $Nb_3Cl_8$ when its Kagome flat band is brought to $E_F$ in a Mott insulating ground state [32]. Therefore, it is crucial to simultaneously bring both Kagome flat band and a dispersive band to $E_F$.

In this article, we experimentally showcase the achievement of a Kagome Kondo lattice in a freshly discovered Kagome material, namely $CsCr_6Sb_6$ (Fig. 1(b)), whose crystal structure is depicted in Fig. 1(c). Unlike other 166-Kagome materials such as $R$$Mn_6Sn_6$ [33] and $R$$V_6Sn_6$ [34] ($R$=rare earth elements) that exhibit a three-dimensional (3D) nature, $CsCr_6Sb_6$ is van der Waals like, featuring a space group of $R$-3m and lattice parameters $a$ = 5.546(2) Å and $c$ = 34.52(1) Å. Figure 1(d) illustrates the building blocks of the double Kagome (DK) layer. Successive three DK layers are shifted through [2/3, 1/3, 1/3] to form an ABC stacking lattice, each segregated by Cs atoms. Figure 1(e) presents an atomic-resolution scanning transmission electron microscopy (STEM) image with the superimposed crystal structure along the $a$-axis. $CsCr_6Sb_6$ single crystals can be mechanically exfoliated into a single DK layer (~1.1 nm, namely 1/3 of the unit cell) as detailed in Fig. S4. Additionally, we note that both single crystals and thin flakes of $CsCr_6Sb_6$ are exceptional robust against moisture, greatly facilitating future characterization and manipulation.

**A Kagome Kondo lattice**

Figure 1(f) presents the angle-resolved photoemission spectroscopy (ARPES) band structures along high-symmetry directions at 15 K. Notably, a dispersionless flat band is observed along the Γ-M, Γ-K and K-M-K high symmetry directions in momentum space, positioned right beneath $E_F$. Enlarged second derivate of band structures can be found in Fig. S2. To further highlight the flat band feature, we plot the energy distribution curves (EDCs) within a narrower energy window in Fig. 1(g). By tracing the EDCs peaks, as indicated by the red tips, the detailed band structure can be discerned. We find that the observed flat band feature is not a single band but consists of two bands around the Γ point. One band crosses the $E_F$, forming a small pocket around Brillouin zone center (as show in Fig. 1(h)), while the other is a dispersionless flat band situated around 40 meV below $E_F$, extending towards the K and M points. Around the K point, the flat band crosses $E_F$, forming a triangular pocket, while near the M point, the flat band just touches $E_F$, creating a prominent spot feature on the Fermi surface.

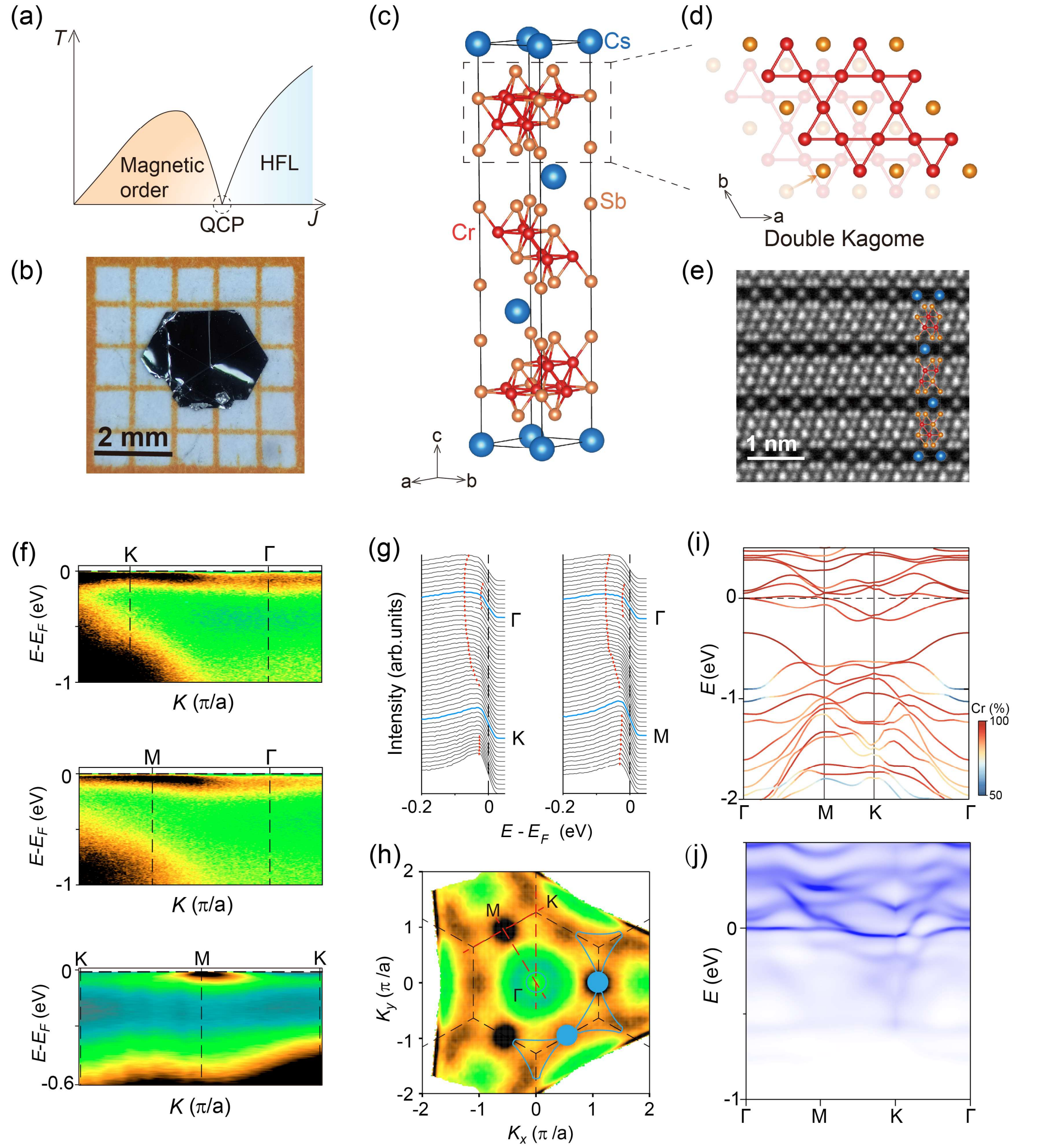

FIG. 1. Crystal and electronic structure of $CsCr_6Sb_6$. (a) Illustration of the competing phases of heavy fermion liquid and magnetic ordered state. (b) Optical image of a $CsCr_6Sb_6$ single crystal. (c) Crystal structure obtained from single crystal X-ray diffraction. (d) Double Kagome structure viewed along the *c*-axis, with the bottom Kagome layer shown in varying transparency. (e) Atomic-resolution image of $CsCr_6Sb_6$ viewed along the *a*-axis with the crystal structure superimposed. (f) Detailed band structures measured along the Γ-K, Γ-M and K-M-K high symmetry directions in the momentum space at 15 K using He I lamp (21.2 eV). (g) Energy distribution curve (EDCs) of the photoemission spectra along the Γ- K and Γ-M high symmetry directions. The EDCs at K and M points are highlighted. The red tips mark the EDC peaks to trace the band dispersions. (h) Fermi surface mapping of $CsCr_6Sb_6$ measured at 15 K. It is obtained by integrating the spectral intensity within 10 meV with respect to the $E_F$. The blue solid curves schematically depict the shape of Fermi surface. (i) Density functional theory (DFT) calculation of $CsCr_6Sb_6$ without considering electron correlation. Color variation represents the spectral weight percentage of Cr relative to the summation of Cr and Sb content. (j) Dynamical mean field theory (DMFT) result of a two-orbital model of $CsCr_6Sb_6$ with $U$=1.8 eV.

Another prominent observation is the presence of a band gap between the flat bands and valence bands, spanning from 50 meV to 0.5 eV. These findings make $CsCr_6Sb_6$ stand out among other Kagome materials, where previously reported flat bands are either mixed with other bands, or positioned far from $E_F$.

Figure 1(i) presents a local-density-approximation (LDA) calculation of $CsCr_6Sb_6$ without considering electron correlation, which reveals a Kagome flat band located exactly at $E_F$. However, we notice that the calculated band dispersions do not align well with our ARPES measurements, particularly around the K point, indicating significant correlation within the system. The band renormalization factor is significantly large, approximately 10, when comparing the ARPES experiments with theoretical calculations. To address this discrepancy, we employ dynamical mean field theory (DMFT) to renormalize the band structure, which successfully captures the features of the flat band at $E_F$, as shown in Fig. 1(j). Notably, the orbital projection (detailed in Fig. S10) shows that all bands near $E_F$ are all primarily contributed by Cr's 3$d$ electrons, confirming that the observed flat bands are pure Kagome bands. These flat bands suppress the kinetic energy of electrons, effectively localizing the magnetic moments of Cr, while residual pockets around the Γ point support mobile electrons. Together, they fulfill the requirements of a Kagome Kondo lattice.

**Heavy fermions with ultra-low charge carrier density**

Figure 2(a) displays the temperature-dependent resistance of a bulk sample. It exhibits metallic behavior at high temperatures and reaches a minimum. Upon further cooling, the resistance increases, showing a canonical -log$T$ divergence. This resistance minimum arises from enhanced scattering between local moments and itinerant electrons, as described by $\rho = aT^5 + c\rho_0 - c\rho_1 \log T$ (ref 35). Below $T_K$=20 K, the resistance deviates from the -log $T$ dependence and continue to increase with a mild slope, marking the formation of Kondo singlets and indicating the emergence of heavy fermions due to Kondo resonance.

To demonstrate this, we compare the heat capacity of $CsCr_6Sb_6$ with its isostructural counterpart $CsV_6Sb_6$ as shown in Fig 2(b) and (c). The extracted Sommerfeld coefficient $\gamma_{Cr\text{-}166}$=75 mJ·mol$^{-1}$·K$^{-2}$ is three times larger than that of $CsV_6Sb_6$. Note that $CsCr_6Sb_6$ exhibits insulating behavior at low temperatures with an ultra-low carrier density of $10^{19}$ cm$^{-3}$ (as shown in the inset of Fig. 2(a), Hall measurements data in Fig. S3), in contrast to the metallic $CsV_6Sb_6$ with a carrier density of $10^{21}$ cm$^{-3}$. This value is even lower than the rare low-carrier-density Kondo systems [36-38], such as $CeNi_2As_2$ [39]. Calculation using the free electron gas model reveals a large effective mass for $CsCr_6Sb_6$ of $m^*/m_0$ = 60, which is two orders of magnitude larger than that of $CsV_6Sb_6$. This value surpassing most $d$-electron systems on record and reaching the territory of $f$-electron heavy fermion materials [17, 40-46] (Fig 2(d)). Instead of a heavy fermion metal, the insufficient Kondo screening results in insulating-like behavior due to incoherent scattering, similar to observations in low-carrier-density f-electron systems $CeNi_2As_2$ and $NaYbSe_2$ [47].

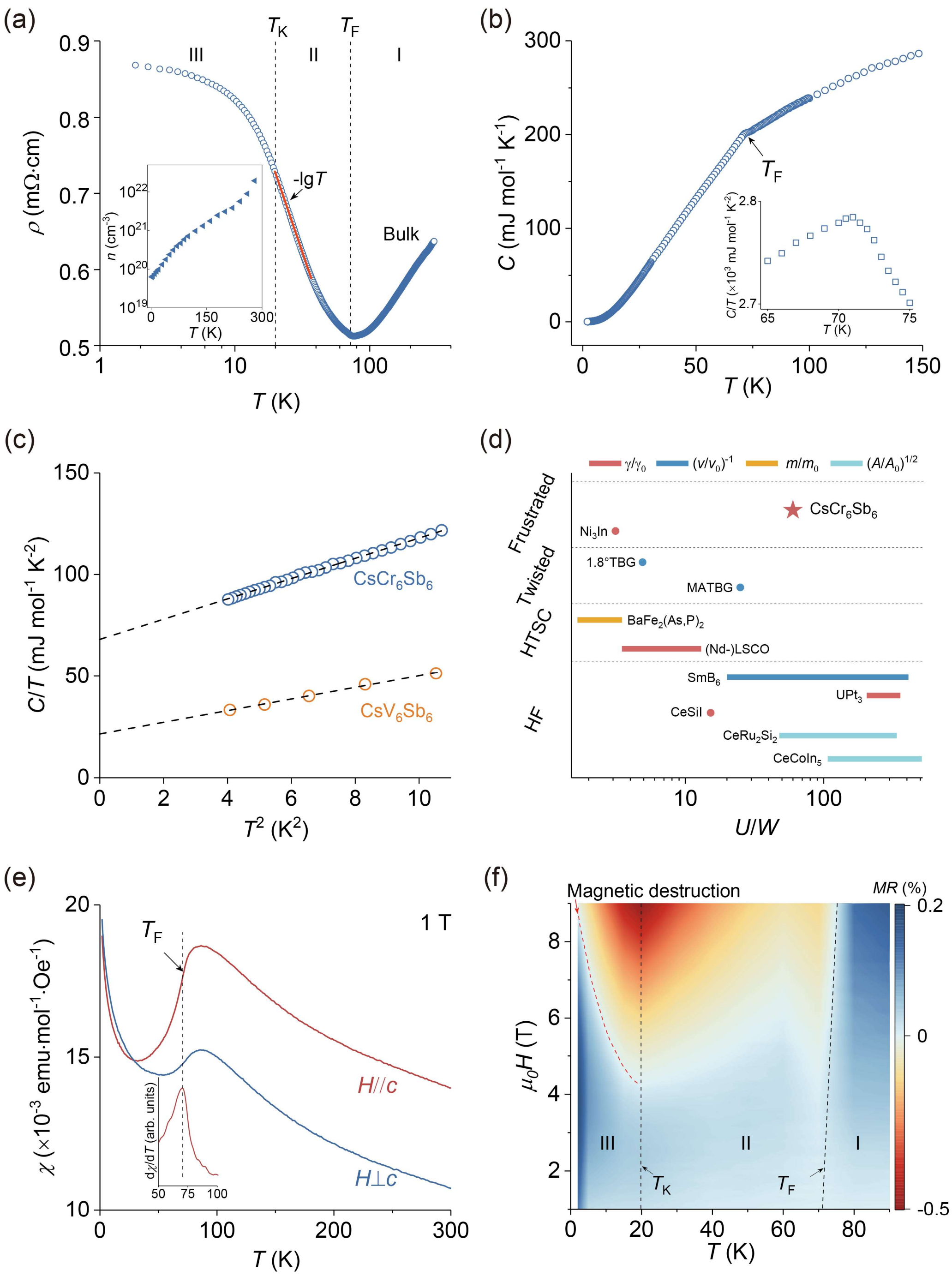

FIG. 2. Frustrated magnetism and heavy fermion with ultra-low carrier density. (a) Resistance of bulk $CsCr_6Sb_6$ single crystal. A kink at 72 K indicates the onset of frustrated magnetic state. The red line is a -log $T$ plot of resistivity, which gradually deviate from linear curve below $T_K$. Inset shows the carrier density extracted from Hall coefficient, which shows a drastic reduction with cooling. (b) Heat capacity ($C_p$) of $CsCr_6Sb_6$ shows a subtle kink at 72 K. Inset is the $T$ dependent $C_p/T$ curve

around 72 K, revealing a broad hump at the same temperature. (c) Linear fitting of $C_p/T$ versus $T^2$, the linear residual gives an electron contribution of 75 mJ mol$^{-1}$ K$^{-2}$. (d) Correlation strength of several correlated materials, which places $CsCr_6Sb_6$ within the realm of *f*-electron heavy fermions [17,40-46]. HTSC and HF are short for high temperature superconductors and heavy fermion materials, respectively. (e) Magnetic susceptibility $\chi$ of $CsCr_6Sb_6$ measured along and perpendicular to the *c*-axis. The d$\chi$/dT curve shown in the inset peaks at 72 K, consistent with the observed $T_F$ shown in (a). (f) Contour plot of *MR,* divided into three regions. Region II and III are dominated by moments and singlets scattering, respectively. Dashed line in region III marks the boundary of magnetic destruction.

Notably, the resistivity exhibits an inconspicuous kink at $T_F$ = 72 K, which is more clearly visible in the derivative shown in Fig. S3. This kink is related to a magnetic interaction, as indicated by the magnetization drop shown in Fig. 2(e). Even though, no long-range magnetic order emerges due to the frustration inherent in the Kagome lattice, as evidenced by the broad hump in magnetization and the absence of the λ-type anomaly in heat capacity shown in Fig. 2(b). The origin of these short-range magnetic interactions remains unclear, potentially arising from the incomplete localization of wavefunction.

Next, we examine the impact of local moment screening on magnetotransport properties. Figure 2(f) shows the contour plot of magnetoresistance (*MR*) of the bulk sample at different temperatures. At $T_F$, a crossover from positive to negative *MR* is manifested, and its amplitude reaches a maximum at $T_K$. These two characteristic temperatures divide the behavior of *MR* into three regions. In region II, the negative *MR* can be attributed to the suppression of local moment scattering by external magnetic fields. In the III (below $T_K$), the partial Kondo screening cancels out the negative *MR* caused by local moment scattering. The *MR* returns to negative at high magnetic fields, which can be attribute to the destruction of heavy fermions and the resurgence of the local moment scattering [48]. Magnetic destruction occurs when the Zeeman energy exceeds $T_K$, so the destruction field increases with cooling. The temperature-dependent *MR* behavior and magnetic destruction of heavy fermions observed in $CsCr_6Sb_6$, combined with the prominent enhancement of effective mass, verify the formation of Kondo resonance at $T_K$.

**Dimensionality induced Kondo breakdown**

The van der Waals nature of $CsCr_6Sb_6$ enables the exploration of dimensional reduction effects on the interplay of various emergent quantum phenomena, which is a long-standing aspiration in study of Kondo lattice physics but remains a significant challenge due to the 3D atomic and electronic structures of conventional heavy fermion compounds [46, 49, 50]. Recent results in exfoliated CeSiI suggest a small step towards this goal, despite the nearly unchanged Kondo temperature and coercive magnetic fields. Here, we demonstrate the successful tuning of $T_K$ in $CsCr_6Sb_6$ by exfoliating the bulk sample approaching its 2D limit.

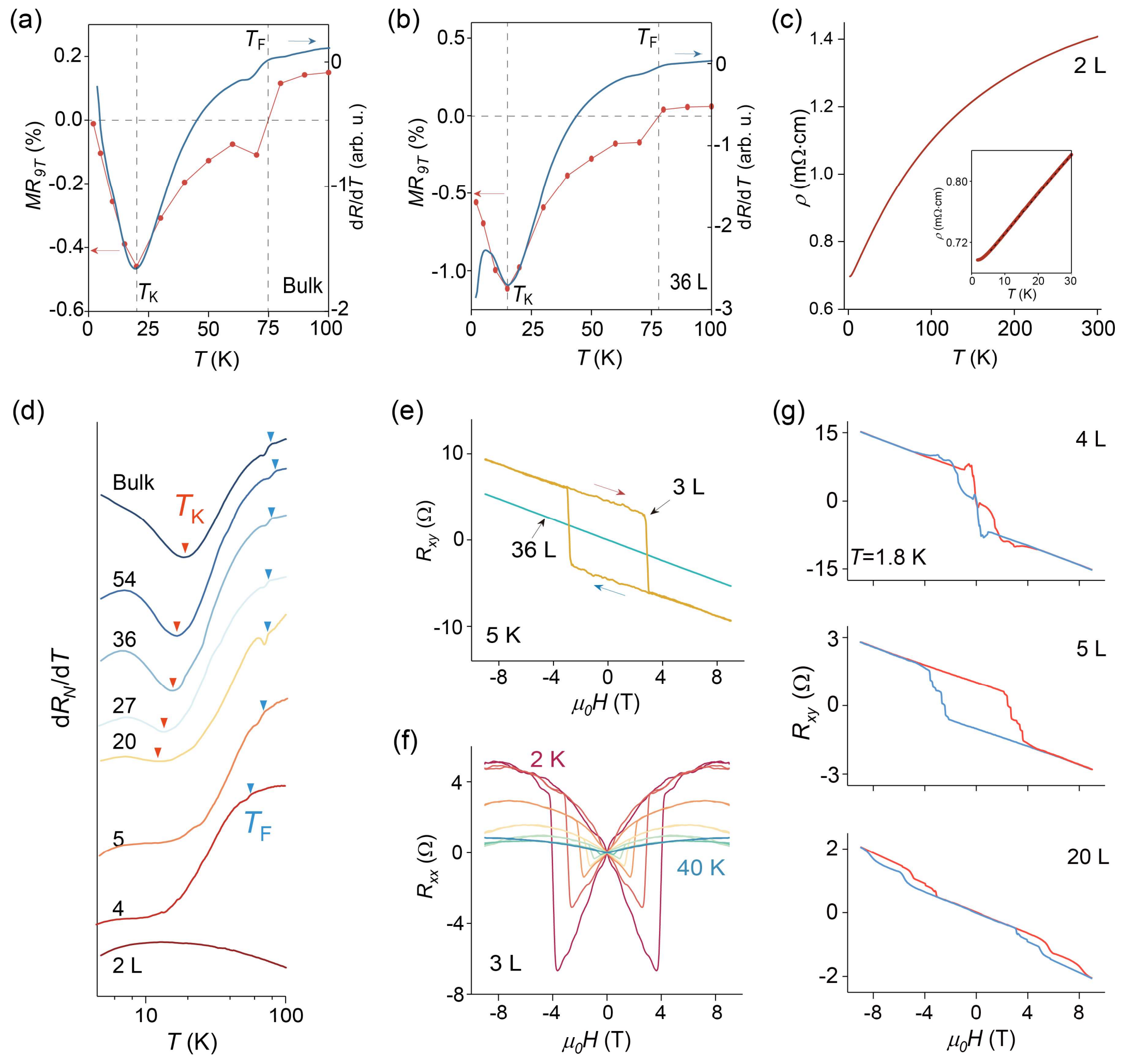

FIG. 3. Highly tunable $T_K$ and a hidden AFM state. (a) and (b) $MR$ at 9 T and d$R$/d$T$ of the bulk and the 36 L sample, respectively. Both bulk and thin flakes exhibit a kink and a minimum in d$R$/d$T$ coinciding with $T_F$ and $T_K$. (c) Temperature dependent resistivity of 2 L sample, with residual resistivity of 700 μΩ·cm. Inset show the linear dependence of the resistivity at low temperature. (d) d$R$/d$T$ of samples with varying thickness from bulk to 2 DK layers (2 L, hereafter). The curves are normalized and vertically offset for clarity. $T_k$ and $T_K$ are indicated by blue and red triangles, respectively. (e) Hall resistance $R_{xy}$ of 3 and 36 L samples. A hysteresis and anomalous Hall resistance appear in the 3 L sample with a coercive field of 3.8 T. (f) temperature dependent $MR$ of the 3 L sample, where the critical temperature is defined as the temperature at which the hysteresis vanishes. (f) $R_{xy}$ of samples with thickness of 4, 5 and 20 L. Even-layer samples exhibit AFM states, while odd-layer ones are in a ferromagnetic (FM) state.

Note that $T_F$ is defined as the temperature at which the $MR$ reverses sign, while $T_K$ is at the maximum negative $MR$. Figure 3(a) and (b) shows $MR$ at 9 T and d$R$/d$T$ of $CsCr_6Sb_6$ in bulk and 36 L sample, respectively. The high similarity of $MR$ and d$R$/d$T$ in both bulk and thin flakes indicates the two characteristic temperatures are both reflected in the resistance behavior. Based on this, we depict the dimensionality effect on the Kondo exchange ($J_K$) in $CsCr_6Sb_6$ by fabricating devices with thickness varies from bulk to 2 L. The d*R*/d*T* data are renormalized and shifted for clarity. We observe that $T_K$ can be feasibly tuned as the sample thickness is reduced to tens of nanometers, exhibiting a monotonic decrease. These results suggest that dimensional reduction diminishes the $J_K$, which is consistent with the general recognition of weakened Coulomb screening approaching the 2D limit. The hopping of conduction electrons is thus suppressed, exacerbating the insufficient Kondo screening. Meanwhile the quantum fluctuations raised by dimensional reduction may also destabilize the singlets.

In the 36 L sample, *MR* measurement indicates that the magnetic destruction field is 2.7 T at 1.8 K (Fig. S5(d)), significantly lower than the 8.7 T observed in the bulk (Fig. 2(f)), verifying the reduced energy scale of $J_K$. By further reducing the thickness to 5 L, the minimum of d*R*/dT vanishes, indicating a complete suppression of $J_K$ and the realization of a dimensionality-induced Kondo breakdown. Meanwhile, $T_F$ persists and continues to shift toward lower temperatures. Interestingly, in the region of complete Kondo breakdown, the 2L sample exhibits strange metal behavior with a residual resistivity exceeding 700 μΩ·cm (Fig. 3(c)), signaling proximity to quantum criticality and a crossover from heavy fermion behavior to strongly correlated Fermi sea.

**Hidden magnetic ordered states**

In a Kondo lattice, the Ruderman–Kittel–Kasuya–Yosida (RKKY) interaction is generally suppressed by Kondo screening. A key distinction between Kondo lattices and impurity Kondo effect is the realization of long-range magnetic ordered states by tuning various parameters. As we have established the dimensionality tuned $J_K$, we further investigate the magnetic response of this Kagome Kondo lattice, particularly near the Kondo breakdown, through magneto-transport measurements.

As expected, we observe the emergence of a long-range antiferromagnetic ordered state below ~20 L, which is absent in the bulk. For comparison, we plot the transverse resistance $R_{xy}$ of 3 L and 36 L thin flakes at 5 K in Fig. 3(e). Distinct from the 36 L and bulk samples, $R_{xy}$ of 3 L sample shows prominent jumps at 2.5 T and display hysteresis upon sweeping the magnetic fields. Notably, the anomalous Hall resistance observed at zero field indicates the formation of a ferromagnetic ground state with the easy axis oriented out-of-plane (Fig. S7). The temperature dependent *MR* of the 3 L sample shows the characteristic butterfly hysteresis of ferromagnetism (Fig. 3(f)).

The odd and even DK layers provide further insights into the magnetic structure. As shown in Fig. 3(g), the even-layer samples remain antiferromagnetic, while the odd-layer samples exhibit ferromagnetic behavior, hallmark of A-type antiferromagnetic ground state. Consequently, the magnetic unit cell of $CsCr_6Sb_6$ should contain at least six DK layers. Meanwhile, the magnetic ordering within each DL and its thickness evolution remains unclear thus far. Moreover, the $R_{xy}$ of few-layers $CsCr_6Sb_6$ flakes show a lot of steps at lower temperatures (additional temperature-dependent $R_{xy}$ data are presented in Fig. S8), suggesting more complex magnetic states or domain structures compared to those observed in $MnBi_2Te_4$ [51]and $MnBi_4Te_7$ [52], warranting further investigation.

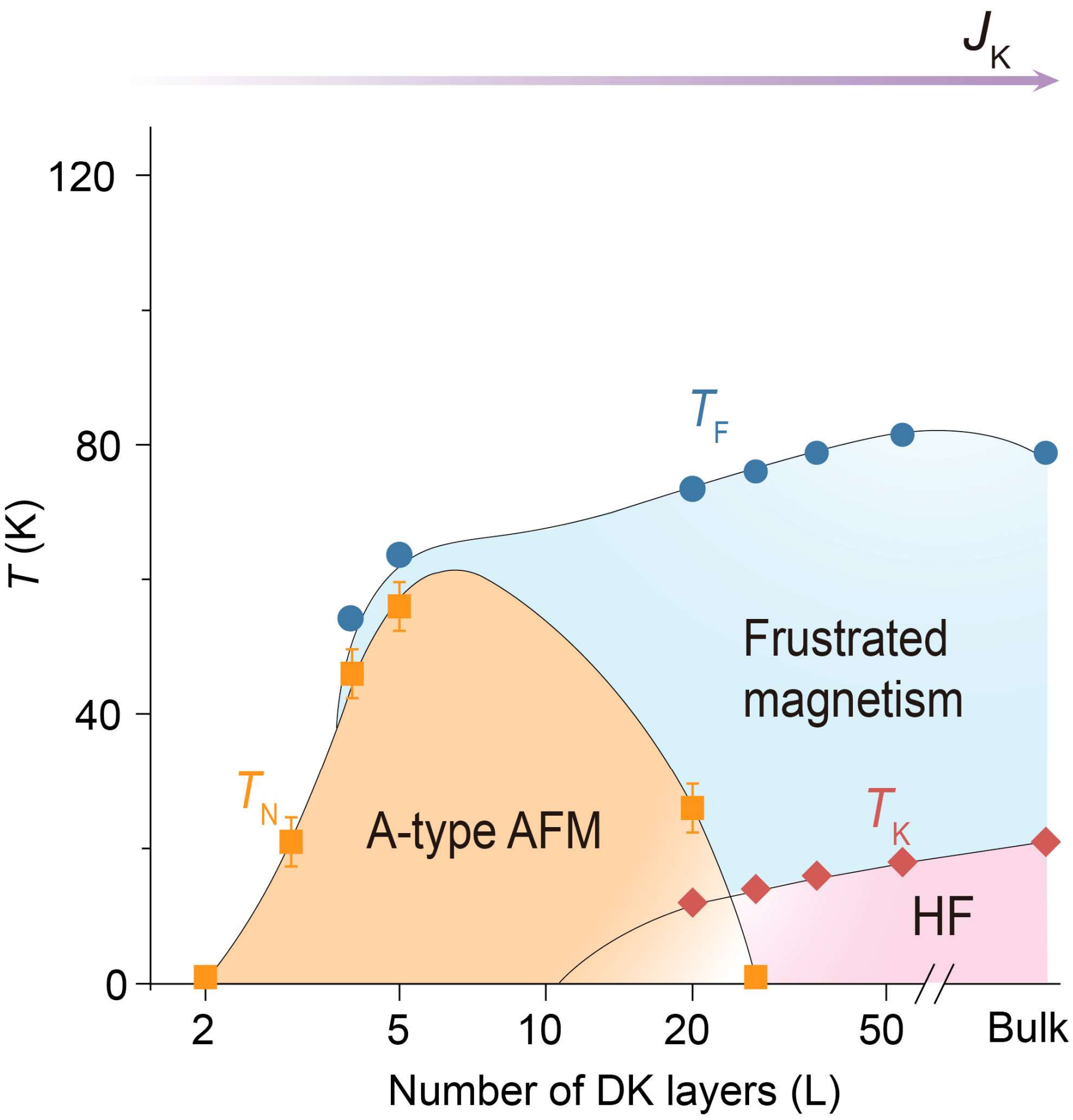

FIG. 4. Magnetic phase diagram. Blue circle and pink diamonds are $T_F$ and $T_K$, respectively. Squares represent the AFM magnetic transition $T_N$, which is determined as the temperature at which the magnetization-induced jumps in $M_R$ or $R_{xy}$ vanish. The thickness of the samples tunes $J_K$ driving the system from a frustrated state in the bulk into a long-range AFM ordered state with dimension reduction down to few layers.

This finding is remarkable from the perspective of 2D magnetism. Recently search for 2D magnetic materials have focused on van der Waals compounds with high Curie or Néel temperatures and strong magnetocrystalline anisotropy to maintain magnetic order approaching their monolayer limit (53-55). $CsCr_6Sb_6$ is the only known example of emergent magnetism upon thinning. This peculiar thickness-dependent magnetism can now be fully understood within a unified framework of Kondo phase diagram.

As shown in Fig. 4, the blue circles and pink diamonds are $T_F$ and $T_K$, and the orange squares represent the onset of magnetic ordering at $T_N$, determined as the temperature at which the magnetization-induced jumps in $M_R$ or $R_{xy}$ vanish. In the bulk, the strong geometric frustration of the Kagome lattice prevents the formation of long-range antiferromagnetic order, which may promote the emergence of the heavy fermions; otherwise, the ground states of the bulk would enter the AFM phase. In the 20 L sample, the emergence of ordered states cannot totally suppress $T_K$. The Kondo screening occurring in an AFM state extends beyond the Doniach phase diagram, where typically only one phase dominants. The exceptional moisture stability of few-layer $CsCr_6Sb_6$ further underscores its potential as a leading candidate for next-generation spintronic devices.

In conclusion, Kagome Kondo lattice has been experimentally realized in $CsCr_6Sb_6$. The discovery of heavy fermion behavior, Kondo singlets, Kondo breakdown, frustrated magnetic states, and a hidden AFM ordering establishes $CsCr_6Sb_6$ as an ideal archetype for investigating various strongly-correlated phenomena. Emergent phenomena including quantum critical point, non-Fermi liquid behavior, heavy fermion superconductivity and high-temperature fractional quantum Hall states are waiting to be explored.

**Acknowledgments**
This work is financially supported by the National Key Research and Development Program of China (No. 2021YFA1401800, 2022YFA1403900), National Natural Science Foundation of China (No. 52272267, 52202342, 52250308). This work was supported by the Synergetic Extreme Condition User Facility (SECUF). Chinese Academy of Sciences President's International Fellowship Initiative. Grant No. 2025PD0005. A portion of numerical computations were carried out at the Hefei Advanced Computing Center.